\documentclass[sigconf,authorversion]{acmart}

\usepackage{paralist}
\usepackage[normalem]{ulem}

\RequirePackage{xspace}

\newcommand{\cp}{Copilot\@\xspace}

\RequirePackage{float}
\RequirePackage{graphicx}
\RequirePackage[skins]{tcolorbox}

\usepackage[os=win]{menukeys} 

\usepackage{lipsum}  
\usepackage{booktabs}
\usepackage{multirow}
\usepackage{graphicx}
\usepackage{colortbl}
\usepackage{listings} 
\definecolor{codered}{rgb}{0.6,0,0}
\definecolor{codegreen}{rgb}{0,0.6,0}
\definecolor{codegray}{rgb}{0.5,0.5,0.5}
\definecolor{codepurple}{rgb}{0.58,0,0.82}
\definecolor{backcolour}{rgb}{0.95,0.95,0.92}
\lstdefinestyle{mystyle}{
    backgroundcolor=\color{backcolour},   
    commentstyle=\color{codegreen},
    keywordstyle=\color{magenta},
    numberstyle=\tiny\color{codegray},
    stringstyle=\color{codepurple},
    basicstyle=\ttfamily\footnotesize,
    breakatwhitespace=false,         
    breaklines=true,                 
    captionpos=b,                    
    keepspaces=true,                 
    numbers=left,                    
    numbersep=5pt,                  
    showspaces=false,                
    showstringspaces=false,
    showtabs=false,                  
    tabsize=2
}
\definecolor{light-gray}{gray}{0.9}

\RequirePackage{pifont}

\setcopyright{acmcopyright}
\copyrightyear{2023}
\acmYear{2023}
\acmDOI{}

\acmConference[SIGCSE '23]{SIGCSE '23: ACM Technical Symposium on Computer Science Education}{March 15--18, 2023}{Toronto, Ontario, Canada}
\acmBooktitle{SIGCSE '23: ACM Technical Symposium on Computer Science Education,
  March 15--18, 2023, Toronto, Ontario, Canada}
\acmPrice{15.00}
\acmISBN{978-1-4503-XXXX-X/18/06}

\begin{document}

\title[Conversing with Copilot: Prompt Engineering Using Natural Language]{Conversing with Copilot: Exploring Prompt Engineering for Solving CS1 Problems Using Natural Language}

\author{Paul Denny}
\email{p.denny@auckland.ac.nz}
\orcid{0000-0002-5150-9806}
\affiliation{%
  \institution{University of Auckland}
  \city{Auckland}
  \country{New Zealand}
}

\author{Viraj Kumar}
\email{viraj@iisc.ac.in}
\affiliation{%
  \institution{Indian Institute of Science}
  \city{Bengaluru}
  \country{India}
}

\author{Nasser Giacaman}
\email{n.giacaman@auckland.ac.nz}
\affiliation{%
  \institution{University of Auckland}
  \city{Auckland}
  \country{New Zealand}
}


\begin{abstract}
GitHub Copilot is an artificial intelligence model for automatically generating source code from natural language problem descriptions.  Since June 2022, Copilot has officially been available for free to all students as a plug-in to development environments like Visual Studio Code.  Prior work exploring OpenAI Codex, the underlying model that powers Copilot, has shown it performs well on typical CS1 problems thus raising concerns about the impact it will have on how introductory programming courses are taught.  However, little is known about the types of problems for which Copilot does not perform well, or about the natural language interactions that a student might have with Copilot when resolving errors.  We explore these questions by evaluating the performance of Copilot on a publicly available dataset of 166 programming problems.  We find that it successfully solves around half of these problems on its very first attempt, and that it solves 60\% of the remaining problems using only natural language changes to the problem description.  We argue that this type of prompt engineering, which we believe will become a standard interaction between human and Copilot when it initially fails, is a potentially useful learning activity that promotes computational thinking skills, and is likely to change the nature of code writing skill development.
\end{abstract}



\maketitle

\section{Introduction}  \label{chap:intro}

Recent breakthroughs in deep learning have led to the emergence of transformer language models that exhibit extraordinary performance at generating novel human-like content such as text (e.g., GPT-3 \cite{brown2020gpt3}), images (e.g., DALL-E \cite{ramesh2022dalle}) and source code (e.g., Codex \cite{chen2021codex}).  Producing source code automatically from natural language prompts promises to greatly improve the efficiency of professional developers \cite{shin2021survey}, and is being actively explored by groups such as OpenAI (Codex), Amazon (CodeWhisperer) and Google (AlphaCode).  After less than one year in technical preview, a production version of Codex called Copilot\footnote{\url{https://github.com/features/copilot}} has recently been released as an extension for development environments such as  Visual Studio Code.  This extension is available for free to students, and claims to be their ``AI pair programmer''.  Just how students will adopt and make use of tools like Copilot is unclear \cite{ernst2022ai}, but it seems certain they will play an increasing role inside and outside the classroom.

Very recent work has shown that these code generation models are good at solving simple programming tasks.  For instance, Finnie-Ansley et al. evaluated the performance of OpenAI's Codex on a private repository of CS1 exam questions, finding that roughly half of the questions were solved by Codex on its very first attempt~\cite{finnieansley2022robots}.  

However, very little is known about the types of problems for which these models tend to fail, or about how students will interact with code generation tools when such failures occur.  One hypothesized interaction that seems very likely is that students will learn to modify, or engineer, natural language problem descriptions to guide the model into generating solutions that ``work'' (at least in the sense of passing available test cases).  Indeed, it is well known that language model outputs are very sensitive to their inputs \cite{reynolds2021computation}.  For example, when using Codex to solve probability and statistics problems, engineering the prompt to include explicit hints on the \emph{strategy} for solving a problem is extremely effective \cite{tang2022solving}.   
An interesting open question in computing education is how well this kind of \textit{prompt engineering} \cite{liu2021pre} works in practice, and whether it might be a useful activity for students who are learning to program.  In this paper, we explore the current performance of Copilot and the efficacy 
of prompt engineering guided by the following research questions:

\begin{itemize}
\item[\textbf{RQ1:}]  How well does \cp perform, at the current time, on a public repository of CS1 programming problems?
\item[\textbf{RQ2:}]  To what extent do natural language modifications to the problem description lead to the generation of successful solutions after \cp initially fails?
\item[\textbf{RQ3:}]  What commonalities exist amongst problems that \cp cannot solve, despite reasonable prompt engineering effort?

\end{itemize}

\section{Motivating example}\label{chap:motivation}

CodeCheck\footnote{\url{https://horstmann.com/codecheck/python-questions.html}} is an online autograder that includes a large collection of publicly accessible programming problems.  The goal of CodeCheck is to assist instructors in offering students exercises they can use for self-practice \cite{kumar2018reflections}.
  
Consider the following problem which appears as Problem 5 under the ``Sums, Averages, Products'' category of the ``Lists (Simple Exercises)'' group in the Python programming problem bank.  The problem is displayed below exactly as it appears on the CodeCheck website, with the problem description appearing as a comment within the code editor, the function signature provided, and the statement ``Your code here...'' as a comment prompting the user to enter their solution:

\begin{lstlisting}[language=Python,basicstyle=\small,breaklines=true,postbreak=\mbox{$\hookrightarrow$\space}]
# Given a list of strings and a string s, 
# return the average length of all strings 
# containing s.

def averageContainingStringS(strings, s): 
    # Your code here...
\end{lstlisting}


What might happen if a student used Copilot as their ``AI pair programmer''
here?  We provided the problem description to Copilot (in July 2022, shortly after its 
general availability release), by pasting the comments and function header, excluding the ``Your code here...'' prompt, into a Visual Studio Code editor with the Copilot extension enabled.
The following suggested solution appeared almost instantly: 

\begin{lstlisting}[language=Python,basicstyle=\small,breaklines=true,postbreak=\mbox{$\hookrightarrow$\space}]
# Given a list of strings and a string s, 
# return the average length of all strings 
# containing s.

def averageContainingStringS(strings, s): 
    count = 0
    for string in strings:
        if s in string:
            count += 1
    return len(s) * count / len(strings)
\end{lstlisting}


This is, of course, \emph{incorrect}---the combined length of the matching strings should be divided by their frequency.
The arithmetic above is  somewhat nonsensical.
If the student who used Copilot to generate this solution submitted it to the CodeCheck website for evaluation, they would see that all of the provided test cases fail.  

One way this error could be resolved is for the student to fix the arithmetic in the code manually.  In this case, Copilot has still served a useful purpose---it helped the student most of the way to a solution by producing code that iterates over the list of strings and that counts the number of matching strings correctly.  However a different approach, and one that we envisage as becoming increasingly common, involves modifying the natural language description of the problem and querying Copilot again.  How could the prompt be engineered to help guide Copilot towards generating a correct solution?  The current (incorrect solution) is not computing the lengths of the matching strings.  Consider the following modification to the prompt which adds comment lines 4--6 outlining a simplistic algorithm for solving the problem.  With these modifications in place, Copilot produced the following solution: 

\begin{lstlisting}[language=Python,basicstyle=\small,breaklines=true,postbreak=\mbox{$\hookrightarrow$\space}]
# Given a list of strings and a string s, 
# return the average length of all strings 
# containing s.
# Count the number of strings that contain s
# Also, sum the length of all strings containing s
# Finally, divide the total length by the count

def averageContainingStringS(strings, s): 
    count = 0
    totalLength = 0
    for string in strings:
        if s in string:
            count += 1
            totalLength += len(string)
    return totalLength / count
\end{lstlisting}


The above solution is correct provided the final value of \texttt{count} is positive (the CodeCheck website does not specify, or test for, the return value if none of the strings in \texttt{strings} contain \texttt{s}).  The comments added to lines 4--6 of this example illustrate one potentially valuable pedagogical use of Copilot:  \emph{describing computational steps in natural language} as a way of guiding code generation models.


Although prior work in computing education has shown that tools like Codex (which powers Copilot) perform well on typical CS1 problems, little is known about the types of problems for which they tend to fail.  In addition, there is currently no work exploring prompt engineering as a strategy for students to resolve errors.  We explore both of these ideas using a dataset of publicly accessible problems, thus establishing a baseline for future evaluations of code generation models which we expect will rapidly improve. 



\section{Related Work} \label{chap:lit-review}

Large language models, or foundation models, are deep neural networks trained with self-supervised learning on broad data sets at a very large scale \cite{bommasani2021opportunites}.  These models can then be adapted, or fine-tuned, for application to a wide range of tasks including the generation of natural language, digital images, and source code.  While their ability to generate novel human-like outputs is on the one hand fascinating, their rapidly increasing deployment has caused alarm among some researchers and led to calls for better understanding of their implications and risks \cite{bender2021dangers, tamkin2021understanding}.

GPT-3, released by OpenAI in May 2020, is a groundbreaking large language model that is trained to predict the next token in a text sequence \cite{brown2020gpt3}.  The Codex model is the result of fine-tuning GPT-3 with an enormous amount of code samples---159GB of code from 54 million GitHub repositories \cite{chen2021codex}.  Copilot is a production version of Codex that has been released as an extension for development environments like Visual Studio Code.  It became generally available to all developers in June of 2022, at which time GitHub announced it would be free for students\footnote{\url{https://github.blog/2022-06-21-github-copilot-is-generally-available-to-all-developers}}.   The impact on educational practice of such technologies is unknown, with arguments on both sides---highlighting concerns of over-reliance by novices \cite{chen2021codex}, and suggesting that the ability to synthesize code automatically could play a revolutionary role in teaching \cite{gulwani2010dimensions}. 

In the computing education literature, there have been very few evaluations to date of code generation models.  
Finnie-Ansley et al. explored the performance of Codex on a private dataset of CS1 exam problems and on several common variations of the well-known rainfall problem \cite {finnieansley2022robots}.   
Codex scored in the 75th percentile when compared to students who were given the same questions, and it was capable of generating multiple correct solutions that varied in both algorithmic approach and code length.  As the complexity of problems grow, it is likely that more human interaction with the models will be needed \cite{austin2021program}.
Sarsa et al. applied Codex to the task of generating novel programming exercises given a single example as input \cite{sarsa2022automatic}.  They found that well over 80\% of the generated exercises included a sample code solution that was executable, but that this code passed the test cases that were also generated by Codex only 30\% of the time. 

Outside of computing education, several recent studies have explored
the potential impact of Copilot for developers \cite{dakhel2022github, nguyen2022empirical, imai2022github}.  Barke et al. observed 20 participants, all of whom had prior programming experience, and found that they were most successful using Copilot when they first decomposed the programming task into microtasks and then prompted Copilot explictly for each of these smaller well-defined tasks \cite{barke2022grounded}.  In particular, they observed that almost all of their participants wrote natural language comments as prompts to Copilot, effectively rephrasing the problem description in natural language.
A very similar finding was reported by Jiang et al., using a different code generation tool called GenLine, where developers would tend to rewrite the natural language problem specifications in order to clarify their intent to the model \cite{jiang2022discovering}.  A similar user study involving Copilot was conducted by Vaithilingam et al. to investigate developer perceptions, interaction patterns and coping strategies when the generated code was not correct \cite{vaithilingam2022expectation}.  They found that when the generated code was incorrect, developers tended to avoid debugging and modifying the code directly, preferring to search for other solutions online or rewrite the code from scratch. 

Research suggests that experienced programmers 
are willing to interact with code generation tools by rewriting problem descriptions in natural language \cite{heyman2021natural}.  We expect this type of interaction to become commonplace as code generation tools are widely adopted, and there is evidence that they perform best when problem-solving strategies and hints are encoded in the prompts \cite{tang2022solving}.  Learning how to effectively converse with code generation tools will therefore likely be an important skill for novices and conversational programmers~\cite{cunningham21conversational} to develop in the near future.

\section{Method} \label{chap:method}

We evaluate the performance of Copilot using Hosrtmann's publicly available CodeCheck exercises\footnote{\url{https://horstmann.com/codecheck}}.  Our evaluation was conducted in July 2022 using the test bank of all `programming problems' available in Python.

\subsection{CodeCheck Python Problems}

The 166 Python problems are split into 23 sub-categories across four main categories (see Table \ref{tab:python-exercises}). The top-level categories are:

\begin{description}
	\item[Branches:] These 22 problems required some combination of \texttt{if}/\texttt{elif}/\texttt{else} statements.  The 13 problems in the \textit{Branches Without Functions} sub-category were the only ones across the entire set where CodeCheck did not provide a pre-defined function header.  For consistency, we prompted Copilot with ``\texttt{def}'' to generate functions for these problems as well. 
	\item[Strings:] These 29 problems required the use of loops (over the characters of an input string), string slicing, indexing, and basic string methods (e.g., \texttt{isdigit()}, \texttt{split()}), but without lists or other data structures.
	\item[Lists (Simple Exercises):] These 65 problems involved searching through lists, counting, averaging, adding/removing/swapping elements, and so on.
	\item[Two-Dimentional Arrays:] These 50 problems involved one-dimensional (list) and two-dimensional (list of lists) arrays and required processing some combination of all elements, or corners, borders and diagonals. 
\end{description}

\begin{table}[]
\small
	\centering
	\caption{Python exercises from the CodeCheck website}
	\label{tab:python-exercises}
	\begin{tabular}{llcc}
		\hline
		     &  &  & Problem  \\
		Category                                  & Sub-category  & Shortcode                & count  \\ \hline
        \multirow{2}{*}{\parbox{1.5cm}{Branches}}		
		& Branches Without Functions & BXF    & 13                   \\
		& Branches with Functions  & BWF      & 9                     \\ \hline
		\multirow{6}{*}{\parbox{1.5cm}{Strings}}		
		 & No Loops   & SNL                   & 5                     \\
		 & Comparing Strings   & SCS          & 3                     \\
		 & Finding Substrings   & SFS         & 5                     \\
		 & Words & SW                         & 4                     \\
		 & Numbers in Strings & SNS            & 3                      \\
		 & Other String Operations & SOO       & 9                     \\ \hline
		\multirow{9}{*}{\parbox{1.5cm}{Lists (Simple Exercises)}}		
		 & No loops  & LNL                     & 4                     \\
		 & Filling & LF                       & 5                     \\
		 & Maximum and Minimum & LMM            & 7                     \\
		 & Finding Elements & LFE             & 6                     \\
		 & Counting Elements & LCE             & 10                    \\
		 & Sums, Averages, Products & LSAP      & 14                    \\
		 & Moving or Removing Elements & LMRE   & 6                     \\
		 & Two Answers & LTA                   & 5                    \\
		 & Double Loops & LDL                  & 8                     \\ \hline
		\multirow{5}{*}{\parbox{1.5cm}{Two-Dimensional Arrays}}		
         & No Loops  & TNL                     & 8                     \\
		 & Loops Along a Row or Column & TLRC   & 11                    \\
		 & Looping Over the Entire Array & TLOA & 13                    \\
		 & Looping Over Neighbors & TLON        & 5                    \\
		 & Producing 2D Arrays & TPA           & 7                     \\
		 & Complex Loops & TCL                 & 6                     \\ \hline
		 & & \textbf{Total}                  & \textbf{166}       \\ \hline
	\end{tabular}
\end{table}

\subsection{Using Copilot}\label{chap:usingcp}

\begin{figure}[btp]
  \centering
  \includegraphics[width=\linewidth]{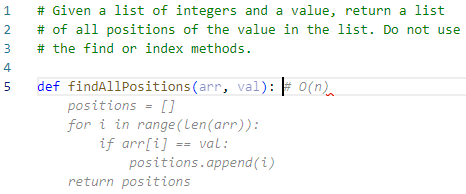}
\caption{Visual Studio Code editor immediately after pasting a CodeCheck problem.  The suggested solution by Copilot appears to the right of the cursor position (to the right of the colon), and can be accepted using the tab key. 
In this example, the suggested solution begins with an in-line comment showing the computational complexity of the code. }
\label{fig:copilotA}
\end{figure}

We copied each problem description from CodeCheck and pasted it into a Visual Studio Code editor with the Copilot extension enabled, as illustrated in Figure \ref{fig:copilotA}.  The 166 exercises in Table \ref{tab:python-exercises} were divided among two of the authors, who used the following protocol:



\begin{enumerate}
	\item \label{itm:start} Copy the problem description (excluding any ``\textit{\# Your code here...}'' comment) and paste it into Visual Studio Code in a blank Python file. 
	\item Wait for \cp to generate a suggestion, and accept it by pressing \keys{\tab} (the \textit{tab} key). 
	\item Select and copy the suggested code from the Visual Studio Code editor and paste  this into the CodeCheck editor.
	\item Press the \menu{CodeCheck} button and record the number of test cases that pass and fail. 
	\item If all test cases pass, declare the problem ``solved'' and move to step~\ref{itm:start} of the next problem.
	\item  \label{itm:delete-buggy-code} If any test cases fail, delete the ``buggy'' code that was suggested by \cp.
	\item  \label{itm:revise-prompt} Observe the failing test cases and engineer the description by 
	adding comments to it that clarify the problem or that provide a strategy for solving the problem.  Do not modify any code---only provide natural language descriptions.
	Repeat steps \ref{itm:delete-buggy-code}--\ref{itm:revise-prompt} until all test cases pass, or there are no obvious clarifications that can be made to the description. 
	\item Record the final engineered problem description, and how many test cases passed as a result of the prompt engineering. Move on to the next problem with step \ref{itm:start}.
\end{enumerate}

Step \ref{itm:revise-prompt} of the protocol was not tightly specified and provided freedom to investigate various prompting strategies.  We considered this appropriate given the exploratory nature of this study, and our goal of identifying the types of approaches that lead to success.  We illustrate typical approaches in Section \ref{sec:prompt_engineering_results}.

\subsection{Categories of Copilot Failures}


Our third research question explores commonalities among the problems that Copilot failed to solve, even after the prompt engineering described in our protocol.
To answer this question, the author who did not use Copilot (Section~\ref{chap:usingcp}) independently reviewed the 34 problems on which which Copilot failed, and categorized them based on the \emph{original prompt}, the \emph{engineered prompt}, and their own (manual) \emph{solution}. Each category suggests a \emph{possible} cause for failures within that category, but establishing definitive causes is outside the scope of this work.

We categorize 15 problems as \emph{Conceptual} (Table~\ref{tab:copilot-failures}), where the original prompt contains one of 6 specific concepts (or terms) that were either retained or slightly reworded in the engineered prompt.
\cp failed on \emph{every} problem (in the full 166-problem set) involving these concepts. For instance, \cp failed on every problem where the \emph{immediate neighbors} of an element in a string or list have to be swapped (two examples are shown in Figure~\ref{fig:failure-swap}), although it successfully solved problems involving other types of swaps. We hypothesize that our failure to ``unpack'' these 6 concepts explains \cp's failure.

Next, we identified instances where our manual solution required a specific code segment to pass all test cases, but no part of the original or engineered prompts corresponds to this segment (e.g., handling ``degenerate'' 2D arrays which have just one row). We hypothesize that this lack of correspondence explains \cp's failure, and we categorize 4 such problems as \emph{Poor Prompts}.


Some of our engineered prompts unpacked concepts in detail (e.g., how to interpret the chessboard position ``b8'' as row and column numbers). \cp is based on Codex, which struggles to parse long prompts~\cite{chen2021codex}. We categorize 11 problems as \emph{Verbose Prompts}, where the length of the engineered prompt appears to be the most reasonable explanation for \cp's failure.

For the remaining 4 problems, we identified ambiguities in the original prompt that allow multiple interpretations. For instance, does reversing the diagonals of a square matrix mean exchanging their elements column-wise, or treating each diagonal as a list to be reversed? In each of these 4 instances, we either failed to address the ambiguity or compounded it in our engineered prompts. We categorize these 4 problems as \emph{Ambiguous}.

Where multiple categories seemed applicable, we preferred the first two categories, since they are backed by evidence: a concept, or a code segment. We provide examples in Section~\ref{chap:commonalities}.

\begin{table}[]
\small
	\centering
	\caption{Copilot failure categories. Subcategory sizes with $\dag$ denote at least one similar problem that \cp could solve.}
	\label{tab:copilot-failures}
	\begin{tabular}{llc}
	\hline
	& & Problem\\
	Category & Subcategory & Count\\
		\hline
		\multirow{4}{*}{\parbox{1.5cm}{Conceptual}} & Largest subarray (2), swap neighbors (3),&\\ & half of odd-length strings (2),& 15\\
		& 2D arrays with different dim. (2),&\\
		& string prefix (3), centre of 2D array (3) &\\
		\hline

		Poor Prompts & Degenerate 2D arrays (2), Other (2) & 4\\
		\hline

	\multirow{4}{*}{\parbox{1.5cm}{Verbose Prompts}} & Pattern ($2^\dag$), Chessboard ($1^\dag$), Time (2),&\\
	& Position of element in a list ($1^\dag$),& 11\\
	& Move/remove elements in a list (3),&\\
	& Longest subsequence ($1^\dag$), Other (1)\\
		\hline
	
		Ambiguous & Adjacent duplicates ($2^\dag$), Other (2) & 4\\
	  \hline
	\end{tabular}
\end{table}


\section{Results and Discussion} \label{chap:results}

\subsection{Initial Copilot Performance}

Of the 166 problems in the CodeCheck dataset, 79 were solved by Copilot on its first attempt (47.6\% success rate).  Table \ref{tab:summary_solved} summarizes the number of problems in each of the primary categories that were solved successfully or remained unsolved even after modification of the problem description.  The `Verbatim' column tabulates the number of problems that passed all of the test cases when the problem description was initially provided as input to Copilot without any changes.  The `Modified' column shows the number of problems that were successfully solved after manual natural language modification of the problem description. 

\begin{table}[]
\small
\centering
	\caption{Number of problems that were solved (when the prompt was provided as input verbatim, and after modification) and that remained unsolved after modification. }
	\label{tab:summary_solved}
	\begin{tabular}{lcccc}
		\hline
		Category  & Verbatim  & Modified & Unsolved  & Total \\ \hline
		Branches & 4 & 13    & 5  & 22     \\  
		Strings & 13 & 7    & 9 & 29     \\  
		Lists & 38 & 19    & 8 & 65      \\  
		Two-D Arrays & 24 & 14    & 12 & 50      \\   \hline
		\textbf{Total} & \textbf{79} & \textbf{53}    & \textbf{34} & \textbf{166}      \\   \hline
	\end{tabular}
\end{table}

\begin{figure}[htp]
  \centering
  \includegraphics[width=\linewidth]{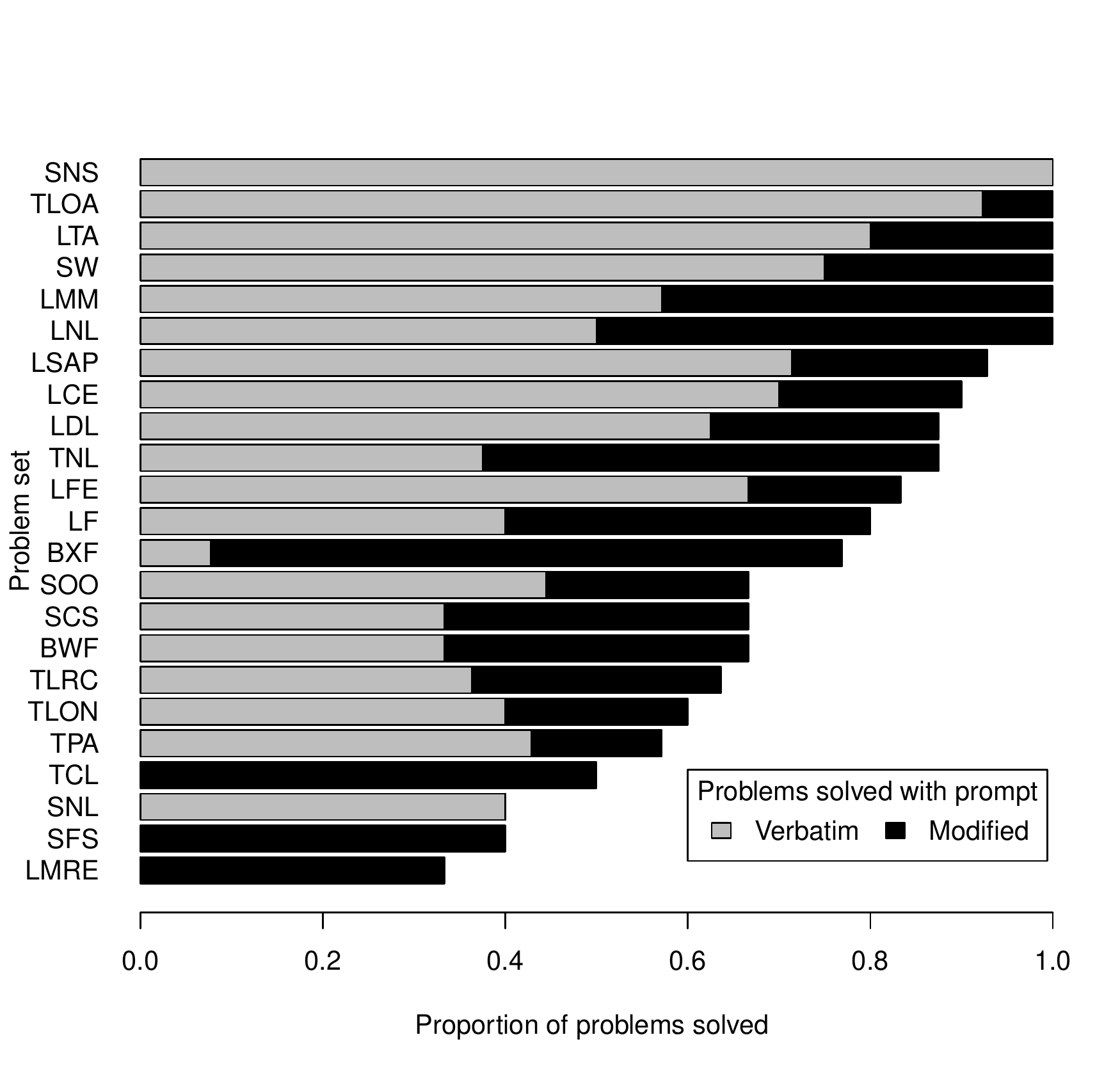}
\caption{Proportion of problems in each sub-category solved using the problem prompt verbatim, and after modifying the problem prompt as outlined in Section \ref{chap:method}}
\label{fig:test_cases_passing}
\end{figure}

\subsection{Effect of Prompt Engineering}
\label{sec:prompt_engineering_results}
A total of 87 problems remained unsolved after the initial code generated by Copilot.  Modifying the description for these problems led to Copilot generating a successful solution in 53 cases (60.9\%).  Figure~\ref{fig:test_cases_passing} further breaks down the performance of Copilot into the problem sub-categories.  Each row corresponds to one of the problem sub-categories, and illustrates the proportion of problems in that sub-category that were solved successfully without any changes to the problem prompt, alongside the additional proportion that were solved after the prompt was modified. 

Performance varied widely across the categories, and although some of this variation is due to categories with relatively small numbers of problems, it is also a result of several categories containing problems with similar solutions.  For example, the SNS category contained three problems that required locating numbers in a string (where words and numbers were separated by spaces).  The same underlying approach, first splitting the string and then iterating over the items, worked for each problem.  

\begin{figure}[htp]
  \centering
  \includegraphics[width=\linewidth]{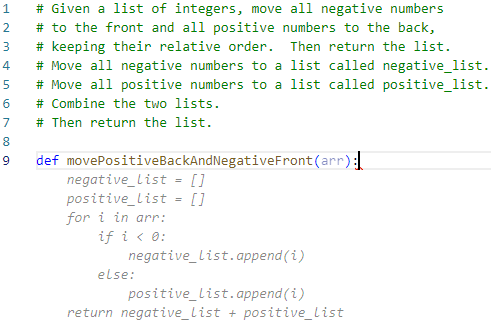}
\caption{Suggested solution (correct after prompt engineering) to problem 4 in the LMRE category.  The manually added comments appear on lines 4--7.}
\label{fig:copilotLMRE}
\end{figure}

Overall performance was worst for the LMRE category, which involved manipulating list elements.  Only two of the six problems in this category were solved by Copilot, and both required prompt engineering.  Figure \ref{fig:copilotLMRE} illustrates one of these two problems, with the original description shown on lines 1--3.  The initial solution generated by Copilot swapped only the first positive and first negative number.  
Prompt engineering (see comment lines 4--7) that explicitly suggested building two lists was successful.  This example highlights that `correctness' is defined by the test cases, and we declare a problem to be solved if all of the CodeCheck tests pass---that was the case for this solution, even though the value 0 is treated as positive.  A more comprehensive set of tests may have necessitated a different solution, and thus more prompt engineering.  Overall, CodeCheck problems had an average of 5.1 test cases per problem.

Prompt engineering was successful to some extent in every category where there was an initial failure with the exception of SNL.  The three problems in this category which consistently failed, despite prompt engineering, involved arbitrary string manipulations and specific special cases.  For example, the full prompt for problem 5 in this category was as follows:  \emph{``\# Given a string, return the string with the first half and the second half doubled. For example, Java becomes JaJavava and Hello becomes HelHellolo. If the length is odd, like Hello, consider the middle character as part of the first half. If the  string is less than 2 characters long, return the original string''}.

In general, when prompt engineering was required, modifications that resembled pseudocode tended to be the most successful.  For example, consider the original description for problem 6 in the TLOA category, which was the only problem in this category that required prompt engineering: \emph{``\#Given a two-dimensional array of integers, shift each row by one to the right and put a 0 at the leftmost column. The rightmost column is lost. Then return the updated array''}.  Copilot tended to wrap the rightmost element into the leftmost position, rather than inserting 0.  The following modified prompt, which is more explicit about the insertion of 0, yielded a correct solution: \emph{``\#For each row: 1) remove the value at the rightmost position. 2) insert the value 0 into index position 0''}.

\subsection{When Prompt Engineering Fails}\label{chap:commonalities}

\begin{figure}[btp]
  \centering
  \includegraphics[width=\linewidth]{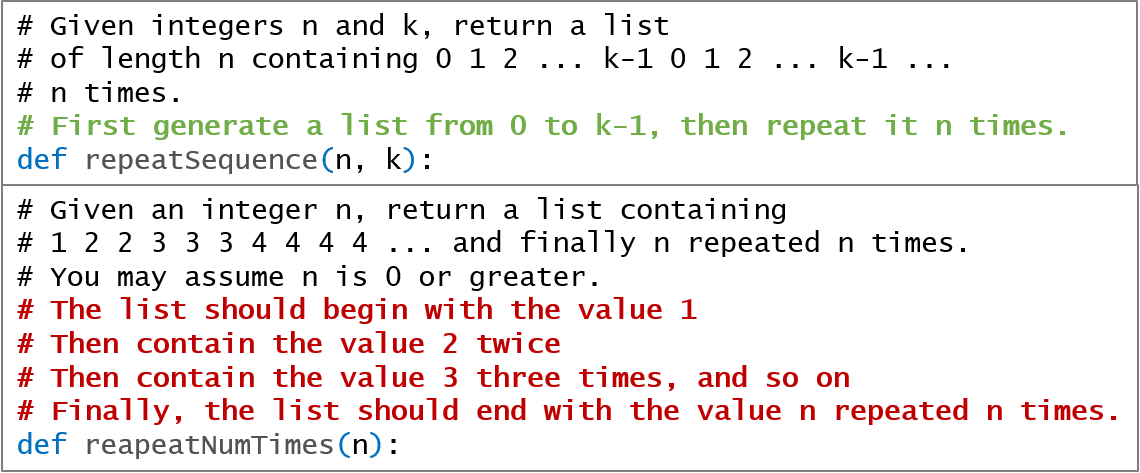}
\caption{\cp solved the upper problem using a terse additional prompt (shown in green), but could not solve the similar lower problem in the \emph{Verbose Prompts} (\emph{Pattern}) subcategory. Note the verbose additional prompt (shown in red).}
\label{fig:failure-verbose}
\end{figure}

In our study, the two largest categories of problems on which Copilot failed (Table~\ref{tab:copilot-failures}) are \emph{Conceptual} and \emph{Verbose Prompts}. We provide illustrations of each type. The original prompt for the upper problem in Figure~\ref{fig:failure-verbose} is poor: it explicitly asks for a list of length \texttt{n}, whereas the test cases expect a list of length \texttt{k*n}. The terse additional prompt (shown in green) helps \cp produce the correct answer (despite retaining the original prompt). In contrast, the original prompt for the lower problem in Figure~\ref{fig:failure-verbose} seems unambiguous. Both problems require \cp to inductively infer a pattern, but \cp fails on the lower problem, despite the additional chain of instructions (shown in red). Chen et al. acknowledge an exponential drop in the model's performance as these chains grow in length~\cite{chen2021codex}. Hence, we categorize this problem as \emph{Verbose Prompts} (\emph{Pattern}).

\begin{figure}[btp]
  \centering
  \includegraphics[width=\linewidth]{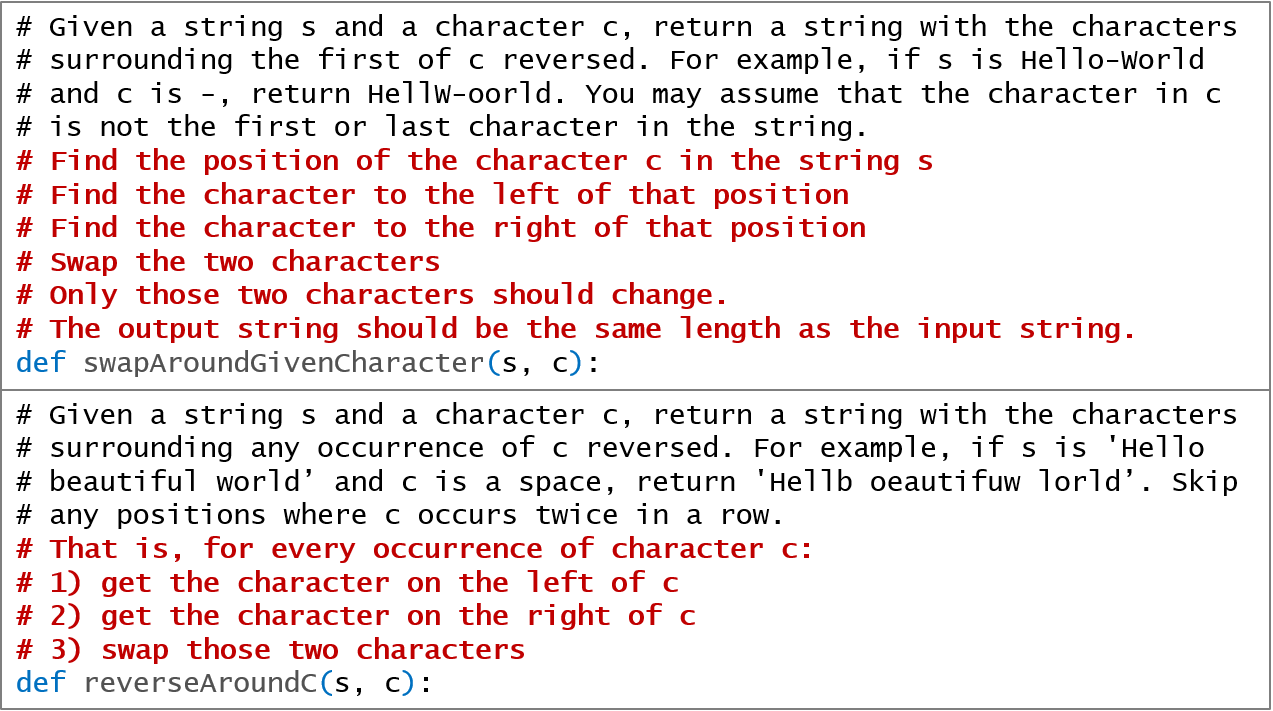}
\caption{Two problems in the subcategory \emph{Conceptual} (\emph{swap neighbors}). The engineered prompts are shown in red.}
\label{fig:failure-swap}
\end{figure}

Figure~\ref{fig:failure-swap} shows two similar problems from the \emph{swap neighbors} subcategory within the  \emph{Conceptual} category. The original prompt of the upper problem specifies that \texttt{c} cannot be the first or last character of \texttt{s}. The lower problem could arguably belong to the \emph{Poor Prompts} category, because neither the original nor the engineered prompt specifies this restriction. Further, for the string \texttt{acbcd}, neither prompt recognizes that the result could be \texttt{bcdca} or \texttt{dcacb} depending on the order in which the swaps around \texttt{c} occur. However, code that resolves these ambiguities is unnecessary since none of CodeCheck's tests examine such inputs. Since the poor prompt does not account for \cp's failure, we categorize this problem as \emph{Conceptual}. In both cases, the engineered prompts are also verbose. This illustrates an interesting tension between the need to unpack concepts while keeping the prompt short. It may be possible to engineer prompts more effectively by rewriting the original prompts, rather than by adding to them as we have done.

\section{Limitations and Future work} \label{chap:limitations}

All problems in this study were in Python, and procedural in nature without utilizing its object-oriented or functional capabilities.  Exploring the performance of Copilot on more complex problems, as well as the use of different prompt engineering strategies, would be valuable to improve our understanding of the use of code generation models in educational contexts.  

The primary focus of our study was on the prompt engineering required to ``converse with \cp''.  Observing novice programmers and their authentic interactions with Copilot would be a fascinating avenue for future work.

Codex, the model that underlies \cp, is non-deterministic \cite{chen2021codex} which introduces challenges around the replicability of our results.  Anecdotally, when returning to some of the problems in our dataset after several weeks, we observed different code suggestions than were originally produced.  Nevertheless, across the 166 problems we evaluated, overall we observe similar rates of initial success reported in earlier studies \cite{finnieansley2022robots}.  As we expect code generation models like Codex to rapidly improve over time, we present our results on a public dataset as a current baseline. 

We see numerous future work opportunities in exploring the impact that \cp will have on programming teaching, learning, and assessment. Open questions remain on the ethics of students using \cp to complete assignments. Is it considered academic misconduct for students to incorporate code suggested by \cp, or is this merely considered an IDE auto-complete feature? How will code similarity tools fare in detecting code written exclusively by \cp?   In what ways can \cp be embraced as a valuable learning tool to help students improve their computational thinking skills?  How will introductory programming courses adapt to the growing use by students of such tools, and how can strategies for constructing effective input prompts be explicitly taught?

\section{Conclusion} \label{chap:conclusions}

Generative language models look set to radically change the way that computing courses are taught and the way that students learn to program.  However, such models are very sensitive to their input prompts, and the ability to engineer effective prompts that generate correct solutions will be an important interaction skill for students in the future.  We present the first exploration of the efficacy of prompt engineering for Copilot in an introductory programming context.  Roughly half of the problems were solved using the original problem descriptions verbatim, and more than half of the remaining problems were solved by engineering the prompts to contain explicit algorithmic hints, which was effective across almost all categories of problems.  We see pedagogical value in these interactions with Copilot, as students need to reflect on code failures and translate the abstract concepts contained in problem descriptions into concrete computational objects and steps, and then express these in natural language.  

\pagebreak

\bibliographystyle{ACM-Reference-Format}
\bibliography{references}

\end{document}